# Maximal benefits and possible detrimental effects of binary decision aids

Joachim Meyer
*Dept. of Industrial Engineering*
*Tel Aviv University*
Tel Aviv, Israel
jmeyer@tau.ac.il

James K. Kuchar
*Massachusetts Institute of Technology*
Cambridge, MA, USA
kuchar@ll.mit.edu

*Abstract*— Binary decision aids, such as alerts, are a simple and widely used form of automation. The formal analysis of a user's task performance with an aid sees the process as the combination of information from two detectors who both receive input about an event and evaluate it. The user's decisions are based on the output of the aid and on the information, the user obtains independently. We present a simple method for computing the maximal benefits a user can derive from a binary aid as a function of the user's and the aid's sensitivities. Combining the user and the aid often adds little to the performance the better detector could achieve alone. Also, if users assign non-optimal weights to the aid, performance may drop dramatically. Thus, the introduction of a valid aid can actually lower detection performance, compared to a more sensitive user working alone. Similarly, adding a user to a system with high sensitivity may lower its performance. System designers need to consider the potential adverse effects of introducing users or aids into systems.

## I. INTRODUCTION

Dynamic warnings, alerts, alarms, and similar decision aids are integral parts in practically all advanced technological systems. These devices can provide important safety information, but they will not necessarily make it easier to control the system. In fact, an abundance of warnings may have negative effects on task performance [1], [2].

So far we lack clear guidelines for deciding under what conditions an aid will improve the overall system performance. The decision to add an aid often seems to derive from the rationale that any additional information will be beneficial, even if its validity is relatively low. This assumption is correct, as long as the human operator assigns optimal weights to the information from the aid and to other available information. However, the benefits from the aid may be very small. If the human operator assigns incorrect weights to the information from the aid, its introduction may in fact lower the overall system performance.

The current paper addresses these issues and aims to provide an evaluation of the possible benefits and costs of introducing an alerting system. We assess the maximal benefits from an alert by computing the maximal detection performance for a combination of a human and a binary decision aid as a function of the human's and the aid's sensitivities. We then present a simple computational method for estimating the combined detection performance when the individual sensitivities are known. We also assess the maximal combined detection performance when the human assigns non-optimal weights to the information from the aid. We show that the availability of an aid or the assignment of part of the responsibility for detections to the human can lower detection performance. Some implications for system design are pointed out.

## II. SIGNAL DETECTION THEORY

We use Signal Detection Theory (SDT) as an analytical framework. It was developed in the 1940s as a formal method for modeling and articulating basic decision tradeoffs [3]-[6]. In the original version of SDT, a decision is to be made that either a signal is present (S1) or not (S0), based upon a single measurement, *y*, corrupted by random noise. The measurement that is obtained is compared against a set threshold value *c*. Depending on whether the measurement is greater or less than the threshold, one of the two hypotheses will be accepted: the signal is not present, or the signal is present. Fundamentals of SDT are well known and we will not discuss them here. The interested reader can find details in the above references.

SDT can be used to examine the tradeoffs in decision outcomes as a function of the placement of the decision threshold. The location of this threshold, c, is often defined in terms of $\beta$, which is the likelihood ratio between the signal-plus-noise ($f_1$) and noise-only ($f_0$) probability distributions measured at the threshold position:

$$\beta = f_{(1)}(c)/f_{(0)}(c) \qquad (1)$$

Fig. 1 shows an example with standard normal distributions, where the strength of the signal, expressed as the distance between the means of the distribution in standard deviations is d'=1. If the threshold is set at *c*= 1, for example, $\beta = 2.7$. Lowering $\beta$ will move the decision threshold left in Fig. 1 and increase the probabilities for true positive, but also false positive responses. Increasing $\beta$ will increase the false negative and lower the false positive probabilities. This tradeoff can be depicted with a Receiver Operating

Characteristic (ROC) curve. ROC curves are plotted along axes of the probabilities of true positives, $P_{TP}$, and false positives, $P_{FP}$ and show the locus of possible operating points as $\beta$ changes. A given selection of $\beta$ maps into a single point on the ROC curve (see Fig. 2). The shape of the ROC curve depends on the probability density function for the noise and the value (or strength) of the signal.

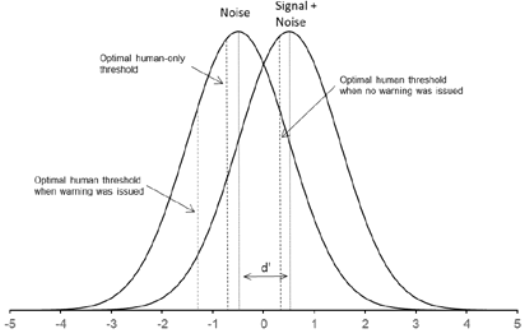

Fig. 1: The basic structure of Signal Detection Theory with two distributions – Noise (S0) and Signal + Noise (S1), the thresholds and the corresponding response criteria β for the human alone and for the human after receiving or not receiving a warning from an alerting system. The distance between the means of the distributions is the sensitivity *d'*.

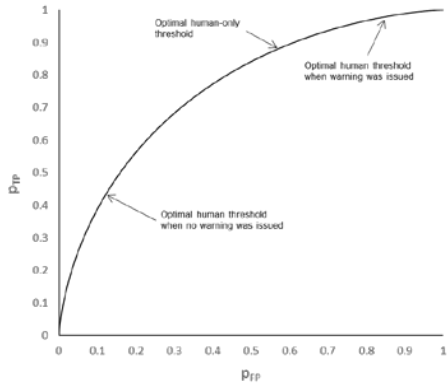

Fig. 2: The ROC curve that corresponds with Fig. 1.

If the costs of various decision outcomes can be defined (e.g., the costs of false positives and false negatives) and the *a priori* probability of the signal is known, then an optimal decision threshold can be set. In particular, if the signal is present with probability $P_{S1}$ and the values of false positives, false negatives, true negatives, and true positives are $J_{FP}$, $J_{FN}$, $J_{TN}$, and $J_{TP}$ respectively, then the optimal threshold setting is

$$\beta^* = \frac{1-P(S1)}{P(S1)} \frac{J_{FP}-J_{TN}}{J_{FN}-J_{TP}} = \frac{1-P(S1)}{P(S1)} U \qquad (2)$$

U is a simplified notation for the payoff ratio. It can also be shown that the slope of the ROC curve at the optimal threshold point is equal to $\beta^*$.

SDT with normal distributions measures a detector's decision quality through *d'*, which is the distance between the means of the two distributions, measured in standard deviations (see Fig. 1). The larger *d'*, the easier it is to distinguish between signals and noise. When the probability density function is Gaussian, and the means of the two distributions are -.5*d'* and .5*d'*, the relationship between $\beta$ and *d'* is

$$\ln\beta = cd' \qquad (3)$$

In more complex cases, the signal and noise distributions may not be known, but observations or predictions of true positive and false positive rates may still be available. These probabilities can be mapped into a single point on an ROC plot. Each point in the ROC plane corresponds to a unique equivalent combination of $\beta$ and *d'*. Thus, given values of $P_{TP}$ and $P_{FP}$, it is possible to calculate the value of *d'* for that system. If $z_{TP}$ and $z_{FP}$ are the inverse of the standard normal distribution values for the $P_{TP}$ and $P_{FP}$ probabilities, respectively, then

$$d' = z_{TP} - z_{FP} \qquad (4)$$

Combining equations 3 and 4

$$ln\beta = -.5(z_{TP}^2 - z_{FP}^2) \qquad (5)$$

When used in this manner, *d'* represents the ability of the system to distinguish the signal from noise as if it were a simple Gaussian SDT problem. Then, *d'* can be used to compare the relative performance of different design settings.

III. SIGNAL DETECTION WITH MULTIPLE DETECTORS

Problems involving more than one decision-maker can be broken down into so-called *tandem* or *parallel* structures. In a tandem structure, a final decision-maker obtains both an independent measurement of *y* and the output of a previous decision-maker. In a parallel structure, several decision-makers each obtain independent measurements of *y*, make separate decisions based on those inputs, and then combine the different decisions in some manner at a fusion center to arrive at a final decision. More complex architectures are also possible, using combinations of tandem and parallel components. General principles of tandem and parallel structures have been investigated to some depth. A study that examined the relative merits of tandem and parallel organizations showed that a tandem structure in which all components make optimal use of the available information is always at least as good as a parallel structure [7].

A particular instance of signal detection in a tandem configuration is the case of a human decision-maker who receives information from an independent decision aid. Such an analysis of alerting systems was provided by [8]. The sensitivity of such a system can be measured through the sensitivity of the combined system $d'_{eff}$. This is the sensitivity equivalent to a single detector with the same level of performance as the combined system. This overall performance is affected by a number of variables

$$d'_{eff} = f(d'_A, d'_H, c_A, c_H, r_{AH}) \qquad (6)$$

where $d'_A$ and $d'_H$ are the sensitivities of the aid and the human, $c_A$ and $c_H$ are the respective cutoffs, and $r_{AH}$ is the correlation between the information that is available to the human and the aid.

For a given prior probability of the signal, a given payoff matrix, and an unaided human's level of discrimination, $d'_H$, there is an optimal cutoff position for the human. If the human also receives information from a binary aid, his or her cutoff also depends on the output of the aid. If the aid indicates that a signal is present, the human should lower the cutoff. That is, the human is more likely to decide that a signal is present because the aid supports that decision. Similarly, if the aid shows that a signal is not present, the human's cutoff should increase.

The optimal difference in the cutoff settings only depends on the diagnostic properties of the aid and is independent of $d'_H$ and of the payoff matrix [9]. Based on (2), the optimal criterion settings when the aid indicates the presence of a signal (A) and when it does not ($\bar{A}$) will be

$$\beta_A^* = U \frac{1-P_{S1|A}}{P_{S1|A}} \quad \text{and} \quad \beta_{\bar{A}}^* = U \frac{1-P_{S1|\bar{A}}}{P_{S1|\bar{A}}} \quad (7)$$

By Bayes theorem

$$\beta_A^* = U \frac{1-P_{S1}}{P_{S1}} \frac{P_{A|S0}}{P_{A|S1}} \quad \text{and} \quad \beta_{\bar{A}}^* = U \frac{1-P_{S1}}{P_{S1}} \frac{1-P_{A|S0}}{1-P_{A|S1}} \quad (8)$$

Fig. 1 presents an example for this adjustment for a situation in which $d'_A = d'_H = 1$, $c_A = -.3$, and the cost of a false positive is one-half the cost of a false negative. Without an aid, the optimal cutoff is at –0.693. Any observed value, $y$, above this cutoff should cause the human to decide that the signal is present. If an aid is placed in tandem, the human has two optimal cutoffs, depending on the output of the aid. If the aid issues an alert, the human's cutoff moves to approximately –1.321; that is, the human is then significantly more likely to agree that the signal is present. If, no alert is issued, the human's cutoff becomes 0.313, and the human is less likely to decide that a signal is present.

The human's decision threshold will depend on whether an alert is present or not, and it depends on the specific threshold setting of the aid. To obtain optimal performance, a coupled relationship between the human and the aid must be solved to find the desired decision thresholds for each decision-maker [10].

[11] describe a tandem structure of decision-makers, using a graphical method for constructing a team ROC curve based on the ROC curves of each of the constituent decision-makers. They also performed an experiment to verify the theoretical behavior, and they did find that team behavior was better than either human or alerting system alone. Additionally, they found data to suggest that humans may operate on two different ROC curves, depending on whether an alert is present. One rationale for this behavior is that humans might pay more attention to a situation in which an alert is present, and therefore be more careful in collecting and interpreting their observations.

## IV. MAXIMAL POSSIBLE $d'_{eff}$

The following analyses are based on a set of simplifying assumptions:

1. The payoff for the operator and the overall system payoff are the same.

2. The costs and benefits of outcomes for the operator are independent of the state of the warning system.

3. The distributions of signal and noise are normal with unity variances, and they are not affected by the appearance of an alert.

4. Events occur independently over time.

5. The information given to the observer and the information on which the aid bases its decision are uncorrelated, given a certain state of the world.

6. Decisions are binary categorizations.

The maximum value of $d'_{eff}$ for a combination of two detectors $m$ and $n$ with $d'_m$ and $d'_n$, respectively, when continuous information is preserved by the detectors and an optimal decision rule is employed, will be according to [12]

$$d'_{max} = \sqrt{d'^2_m + d'^2_n} \quad (9)$$

To assess the maximal $d'_{eff}$, denoted $d'_{eff}*$, we computed values of $d'_{eff}$ over a range of combinations of $d'_A$ and $d'_H$. This was done by determining the optimal operator cutoff for a given alerting cutoff, using (8), computing $z_{FP}$ and $z_{TP}$, and then using (4) to compute $d'_{eff}$. The alerting system cutoff was then systematically varied (with the human cutoff changed in response) until the value for $d'_{eff}$ was maximal.

The dots in Fig. 3 show the values of $d'_{eff}*$ for various combinations of $d'_H$ and $d'_A$. The diagonal shows $d'_{eff}*$ for a human-only condition ($d'_A = 0$). The optimal benefit of the tandem combination of an aid and a human is then given by the distance of each curve above this diagonal. For example, with $d'_A = d'_H = 1$, the best possible combined sensitivity is $d'_{eff}* \approx 1.3$. An aid with significantly poorer performance provides little or no benefit to the human (e.g., with $d'_A = 0.5$ and $d'_H = 1$, $d'_{eff}* \approx 1.08$). Conversely, when an aid is significantly better at discriminating the signal compared to the human, the human is of little added benefit to the system (e.g., with $d'_A = 2$ and $d'_H = 1$, $d'_{eff}* \approx 2.06$). The largest tandem benefits occur when the human and the aid have similar values of $d'$. Note also that all benefits are smaller than predicted by (9), because the information from the aid is binary, rather than continuous.

The method presented above does not lend itself easily to the computation of $d'_{eff}*$ for a system. We developed here an approximation that provides an estimate of $d'_{eff}*$ for given values of $d'_A$ and $d'_H$. This makes it possible to assess the optimal benefit that can be gained from adding an aid or from involving the human operator in a decision that could be automated, based on the information available to the aid.

As a starting point, we assumed that $d'_{eff}*$ will be similar to the combined $d'$ for two information sources (see 9), but with some loss of information due to the aid providing only binary information. To estimate $d'_{eff}$, we used the expression

$$d'_{eff} = \sqrt{d'^2_H + d'^2_A - ad'_H d'_A} \qquad (10)$$

When applying the expression to values depicted in Fig. 2, we obtained the best predictions of $d'_{eff}$ from $d'_H$ and $d'_A$ for a=.3024, for which the computed value predicted 99.7% of the variance of $d'_{eff}$. A reasonable approximation for the maximal combined sensitivity $d'_{eff}$ of a human detector with $d'_H$ and a binary alert with $d'_A$ is

$$d'_{eff} = \sqrt{d'^2_H + d'^2_A - .3 d'_H d'_A} \qquad (11)$$

The correlation between predictions and observed values in this case is .998 and predictions are drawn as lines in Fig. 2.

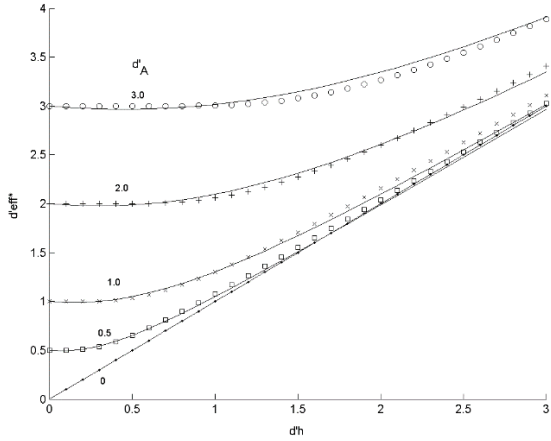

Fig. 3: Maximal possible combined sensitivity values ($d'_{eff}*$) for a human operator with sensitivity $d'_H$ and an aid with $d'_A$. Lines indicate the maximal combined sensitivity $d'_{eff}$ as predicted by the expression $d'_{eff}=(d'^2_H+d'^2_A-.3d'_H d'_A)^{.5}$.

## V. EFFECTS OF NON-OPTIMAL SETTINGS OF $C_H$

The analysis up to now concentrated on the optimal benefits that could be gained from introducing an aid. However, people may fail to combine information from the different sources optimally.

Based on (8), for two cue values A and Ā (alert and no alert) a consistent observer will adopt cue-contingent criteria, which are proportional to the cues' likelihood ratios under signal and no-signal [9].

$$\beta_A : \beta_{\bar{A}} :: \frac{P_{A|S0}}{P_{A|S1}} : \frac{P_{\bar{A}|S0}}{P_{\bar{A}|S1}} \qquad (12)$$

We can therefore define for each system a cue-contingent optimal criterion shift

$$\Delta ln\beta^* = ln\beta_A^* - ln\beta_{\bar{A}}^* \qquad (13)$$

which must be used to obtain maximal benefits from the alerts. The intuition behind this expression is that the optimal criterion shift depends on the validity of the cue. For a perfectly valid aid the observer should always respond S when the cue value is A ($ln\beta_A=-\infty$) and should never respond S when the value is Ā. For an entirely uninformative cue $ln\beta_A = ln\beta_{\bar{A}}$ (i.e., the criterion setting is independent of the information from the aid). For intermediate levels of aid validity, $\Delta ln\beta^*$ depends on the validity of the aid.

In order to determine the effects of non-optimal criterion settings by the operator we computed again $d'_{eff}*$, but this time using a model of varying levels of confidence in the aid. To model the degree of trust the human places in the aid, we varied the ratio between $\Delta ln\beta$ (the difference between the two cutoffs the human uses) and $\Delta ln\beta^*$ (the optimal difference in cutoffs). If this ratio was 1, the human's trust in the aid corresponds to the optimally warranted level of trust. For ratios greater then 1, the human puts excessive trust in the aid, and for ratios smaller then 1, the human does not trust the aid sufficiently.

Fig. 4 shows the resulting combined system performance as a function of the ratio $\Delta ln\beta/\Delta ln\beta^*$ over several values of $d'_A$. Performance is normalized by $d'_H$. With small values of the ratio, performance of the combined system approaches that of the human alone. For large values of the ratio (over-trust) the combined system performance approaches that of the aid system alone. If the ratio is 1, an optimal level of trust is placed in the aid, and the combined system performance ($d'_{eff}/d'_H$) is maximal.

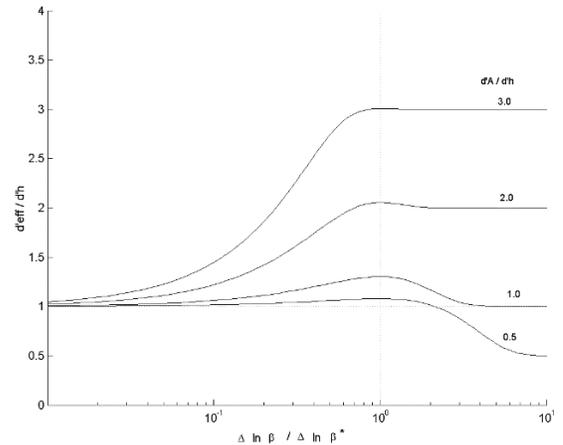

Fig. 4: Effect of level of trust on effective system sensitivity as expressed by the ratio $d'_{eff} / d'_H$ for aids with different levels of validity, compared to the human, as expressed by the ratio $d'_A / d'_H$. The X-axis is the ratio between the observed and the optimal differences between threshold settings with and without alerts.

Fig. 4 also shows some additional points. First, as can be seen in Fig. 3, the largest possible benefit of the combined human + aid system exists when the human and the alert have similar sensitivities ($d'_A/d'_H = 1$). For low-validity aids (e.g., $d'_A/d'_H = 0.5$), the user can gain very little benefit from the aid even at an optimal trust level. Should the human over-trust a low-validity aid, the combined system runs the risk of operating below the human-alone performance. If the human under-trusts a low-validity aid, the combined performance is not otherwise harmed. Similarly, for a high-validity aid (e.g., $d'_A/d'_H = 3$), the combined system performance will be less than that of the aid alone if the human under-trusts the aid. For each level of $d'_A/d'_H$, there is a small window of trust levels for which an overall benefit exists. Trusting too much or too little (going outside this window) results in system performance approaching either the human-alone or the aid-alone performance, reducing the usefulness of the tandem system design.

One implication of Fig. 3 is that when the human and the aid have very different sensitivity levels, there is little benefit to running them in a tandem configuration. More importantly, small shifts in trust can produce significant losses in performance over using the better of the two detectors alone. Designers should therefore make efforts to ensure that operators will not over or under trust a system, especially when the sensitivities are significantly different. A tandem system in which human and alert are of similar sensitivities is not prone to this loss of performance.

## VI. COMPARISON TO EXPERIMENTAL RESULTS

A large body of research on human decision-makers' criterion choice in signal detection problems has shown that the human's cutoff is often not sufficiently adjusted to the prior probabilities of signals and the payoff matrix, a phenomenon named "sluggish beta" [13]. Many early studies on signal detection in psychology showed this effect [14]-[18], as did applied studies on visual inspections [19] or the reading of dental medical imagery [20]. Similar findings were also obtained in quality control inspection tasks [21], in military target detection [22], [23], and in other domains.

The phenomenon also occurred in aided decisions. In an early study of signal detection with an additional binary information source participants had to detect an auditory stimulus and were aided by a visual cue that was supposed to indicate the likely occurrence of the auditory stimulus [24]. Three experiments varied the detectability of the acoustic stimulus ($d'_H$ in our notation), the criterion setting of the cue ($\ln\beta_A$) and the validity of the cue ($d'_A$). Participants used different criteria with and without a cue, but the difference in the criterion settings for the different cue conditions was clearly smaller than optimally required. Thus participants seemed to give too little weight to the cue. One exception was a condition in which the cue was highly valid. Here participants seemed to limit the attention they gave to the auditory stimulus and gave more weight to the cue, in spite of the fact that the auditory stimulus alone was still more detectable than the cue.

In another study, participants had to detect the appearance of a longer line segment among 16 lines that were presented in four groups of 4 lines each [9]. Participants were in some conditions aided by a visual cue. The experimental conditions differed in the specificity of the cue (the cue either indicated the possible existence of a long segment in any group or it indicated the group of lines in which the segment was supposedly located), the diagnostic value of the cue, and the criterion setting of the cue. Overall participants responded to the cue and used different cutoff settings when a positive or a negative cue was given. However, they responded less then optimally warranted and they benefited from the cue only when it was location specific.

Two studies [25], [26] dealt with aided signal detection in a simulated production task where participants had to decide whether to produce or not, based on a continuous stimulus and a binary warning that either showed a red or green light. Results showed that participants adjusted their weighting of the information from the warning according to the diagnostic value of the warning, but this weighting was generally less than optimal. In experimental conditions with non-valid warnings, i.e., when there was no correlation between the warning output and the state of the system, participants continued to respond to the warning cue. These findings also provide evidence for the insufficient adjustment of cutoff settings (i.e., the "sluggish $\beta$") in aided signal detection.

Another study of binary categorization performance reports two experiments with a novel experimental tool, in which participants decided whether items on a screen were intact or faulty, based on the configuration of lighter and darker areas in items [27]. Cues were available in half of the experimental blocks, and participants could use them in their decisions. Experimental conditions differed in the effort required to perform the task, manipulated through the contrast between lighter and darker areas (higher contrast vs. lower contrast), and in the validity of cues (medium vs. high validity). The required effort in the task did not affect the strength of responses to cues. While participants did adjust their responses to the validity of the cues, this adjustment was clearly less, and their responses to the cues were weaker, than would be required according to a normative model.

In terms of our analysis, respondents were always to the left of the $10^0$ X value in Fig. 4 when the alerting system had more than minimal sensitivity. Thus, they did not use the full capability of the alert, and their performance was lower than the performance that could have been achieved if the system would have been entirely automatic. Also, when the alerting system had very low or no sensitivity at all, participants assigned it too much weight, and therefore their performance was lower than could have been achieved without the alert.

## VII. CONCLUSIONS

Our paper analyzes the effect of a binary aid on signal detection performance. The analyses presented here indicate that binary aids may often have only limited value, if users are able to perform the task well without the aids. Furthermore, estimates of the maximal possible benefits from the aid are probably high, considering that they are based on the assumption of uncorrelated information for the observer and the aid, given a certain state of the world. In fact, the aid and the operator will often receive correlated information, which will lower the possible benefits from the aid.

The introduction of an aid may actually lower performance if the user assigns non-optimal weights to the information it provides. This will be the case when the user gives excessive weight to the information from the aid, which is less valid than the information he or she receives independently. In the extreme case, a user may rely on the indications of a non-valid aid, which would make responses uncorrelated to the actual state of affairs.

The introduction of a user into a system with a highly valid aid may also lead to less than optimal performance. If the aid is more valid than the information the operator has independently, the operator may override indications from the aid, even though her or his ability to distinguish between states of the world is smaller than the aid's ability. Any design decision that combines the human and an aid must consider these possibilities.

The empirical results show that people tend to put excessive weight on non-valid aids and give insufficient weight to valid aids. Thus, in many cases, people do not reach optimal performance with a tandem system. Consequently, at least in terms of the detection performance, the better detector alone (be it the human or the aid) could perform better than the incorrect combination of the human and the aid. The benefit from combining the two is often very small, especially when their detection abilities differ.

Some caution is, however, warranted. Aids may provide benefits that go beyond simple signal detection. For instance, a user could potentially use the information from an aid for other purposes that go beyond the signal detection task, such as helping the user to maintain situational awareness. There is indeed evidence that aids fulfill complex functions for experienced operators (e.g., [28]). These functions must be understood when designing the aids in a system.

We also do not consider the possible costs or benefits that are related to aids in terms of the operator's workload. An aid may increase the workload and make it more difficult for the operator to perform her or his task, even if the system may somehow raise the overall sensitivity. Alternatively, the information from an aid may be available with less effort than information from other sources. Consequently, there may be an advantage to having an aid, even if its sensitivity is relatively low, if with the system the operator needs to invest less effort in monitoring the information, and the task therefore becomes easier.

Still, while a signal detection analysis should not be the ultimate criterion for deciding on the installation of an aid in a system, it should be a part of the evaluation of such a device. For this device to have any value, it should either raise detection performance in terms similar to those presented here, or it needs to provide some other benefit that must be clearly specified. If neither of these benefits can be demonstrated, one should carefully consider if one wants to add an aid to a system. A broad modeling of the use and the response to aids is needed to provide designers with guidelines on how and when to implement aids in systems.